\documentclass[12pt]{article}%

\usepackage{psfig,float}
\begin{document}
\setlength{\baselineskip}{24pt}

\begin{center}
{\large{\bf Two complementary representations of a scale-free network}}
\end{center}

\begin{center}
J.C. Nacher\footnote{Corresponding author: E-mail:nacher@kuicr.kyoto-u.ac.jp. Tel: +81 774 383020. Fax: +81 774 383022}, T. Yamada, S. Goto, M. Kanehisa and T. Akutsu
\end{center}

\begin{center}
{\it Bioinformatics Center, Institute for Chemical Research, Kyoto University}
\end{center}

\begin{center}
{\it Uji, 611-0011, Japan}
\end{center}

\vspace{1.0cm}

\begin{abstract}
{\small{ Several studies on real complex networks from different fields as biology, economy, or
sociology have shown that the degree of nodes (number of edges connected to each node) follows a scale-free power-law
distribution like $P(k)\approx k^{-\gamma}$, where $P(k)$ denotes the frequency of the nodes 
that are connected to $k$ other nodes. Here we have carried out a 
study on scale-free networks, where a line graph transformation (i.e., edges in an initial network are transformed into nodes) 
is applied to a power-law distribution. Our 
results indicate that a power-law distribution as $P(k)\approx k^{-\gamma +1}$ is found for the transformed network together with a 
peak for low-degree nodes. In the present work we show a parametrization of this behaviour and 
discuss its application to real networks as metabolic networks, protein-protein interaction network and World Wide Web.}}
\end{abstract}



\section{Introduction}

Commonly, networks of complex systems were described with the classical random graph
theory of {\it Erd$\ddot{o}$s} and {\it R$\acute{e}$nyi} \cite{1,2}. In the random graph theory, the degree distribution $P(k)$ (probability
that a randomly selected node has exactly $k$ edges) peaks strongly 
around $K$ = $\langle k \rangle$, where $\langle\rangle$ denotes the average. However, recent 
experimental studies indicated that the random graph model
could not explain the topological properties of real networks. 
For many real networks the degree distribution was found to follow a scale-free power-law 
distribution like $P(k)\approx k^{-\gamma}$ with an exponent $\gamma$ between two to four. Some examples of these networks are
World Wide Web ($\gamma = 2.1$), power grid ($\gamma=4$) and film actors ($\gamma=2.3$) \cite{7}. The analyses also showed that some physical networks 
(neural network of worm $C.elegans$, film actors, power grid) had common features as
small path length $L_{actual} \approx L_{random}$ but high clustering degree $C_{actual} >> C_{random}$ \cite{3}, revealing a 
different topology than the classical random approach. 
The systems with these properties were called "small-world" networks \cite{4,5,6}.

Concerning the biological networks, several analyses on metabolic networks corresponding to 43 different organisms \cite{8}
and protein-protein interaction network \cite{9} were done. These works revealed the same scale-free topology found 
in non-biological networks.

This wealth of data stimulated to develop
theoretical approaches to reproduce such kind of scale-free topology.
One of the most successful models was proposed by {\it Barab\'{a}si-Albert} \cite{7,10}, which introduced a mean-field method to 
simulate the growth
dynamics of individual nodes in a {\it continuum theory} framework. The {\it Barab$\acute{a}$si-Albert} (BA) model is based on two main
mechanisms; (1) Growth: the network starts with a
small number of nodes ($m_0$), and a new node is added at every time step with
($m\leq m_0$) edges that link the new node to ($m$) different nodes. (2)
Preferential attachment: the BA model assumes that the probability $\prod$
that a new node will be connected to node $i$ depends on the connectivity
$k_i$ of that node $\prod(k_i)$ = ${k_i}/{\sum_j k_j}$. Therefore, after
$t$ time steps, the network is populated with $N$ = $t$ + $m_0$ nodes and $mt$
edges \cite{7,10}. Even though recent extensions of this model, with rewiring edges \cite{11} , adding a 
fitness-dependent dynamic exponent \cite{12}, and
with aging features \cite{13,14} have provided a more accurate description
of the network evolution, by generating a large available spectrum 
of scaling exponent or cut-offs in the connectivity distribution \cite{12}, we will use the original BA model 
for 
generating our synthetic network. 

The aim of our work is 
to introduce the concept of the line graph transformation and to study the topology of 
the scale-free networks after the line graph transformation is done. As the line graph transformation is closely related to
the representation of metabolic networks, similarities and differences between the line graph transformation 
and real metabolic networks are also
discussed. In addition, we will illustrate our results with examples from several real networks as World Wide Web and 
protein-protein interaction networks. 

It is also worth noting that the line graph transformation has recently been applied with success 
on the protein interaction network \cite{protein} with the aim to detect functional modules. In that work, the 
edges (interactions) between two proteins 
become the nodes of the transformed network (interaction network). By means of the line graph transformation, the interaction
network has its structure level more increased than that from the protein network (i.e., higher clustering coefficient), and by 
using the TribeMCL algorithm \cite{protein2} they are able to detect clusters in the more highly 
clustered interaction network. These clusters are transformed back to the initial protein-protein network to identify 
which proteins conform functional clusters. 

The paper 
is organized as follows. In Sec. II we describe the theoretical model that we use and we 
explain in detail the mathematical methods. In Sec. III we present the experimental data of several real networks 
and we compare with our
theoretical predictions. The final section summarizes our work.



\section{Theoretical Models}

\subsection{Line graph transformation}

Given an undirected graph $G$, defined by a set of nodes $V(G)$ and a set of edges $E(G)$, we associate another graph $L(G)$, called
the line graph of $G$, in which $V(L(G))=E(G)$, and where two vertices are adjacent if and only if they have a common endpoint in $G$ (i.e.,
$E(L(G))=\{\{(u,v),(v,w)\}| (u,v) \in E(G), (v,w) \in E(G)\}$). This construction of
graph $L(G)$ from the initial graph $G$ is called line graph transformation \cite{15}. 

In Fig. 1(a), we consider a graph $G$ and we apply the
line graph transformation  over this  graph $G$. The result of this transformation is the line graph of $G$, $L(G)$.
We see that, under the line graph transformation, the nodes of $L(G)$ are the edges of $G$, with two nodes of $L(G)$ adjacent whenever 
the corresponding edges of $G$ are.
If $x=(u,v)$ is an edge of $G$, then the degree of $x$ (deg $(x)$) (i.e., number of edges connected to the node $x$) in $L(G)$ can be 
written as: deg $(u)$ + deg $(v)$ - 2. 
The example shown in Fig. 1(a) can correspond to a synthetic network generated by the BA model, where the
network is only composed by one kind of nodes. In this case, the edges of the initial network $E(G)$ do not correspond to 
any physical entity. The line graph transformation becomes more meaningful when those edges $E(G)$ have physical meaning.   

This concept of line graph transformation has similarities to the representation of metabolic networks as we can see in Fig. 1(b). 
Fig. 1(b) (left) shows a real subgraph from the Lysine Biosynthesis metabolic network in the ordinary representation
of biochemical pathways. We see that the nodes are chemical compounds and edges are represented by chemical reactions \footnote{It is
important to comment the differences between enzymes and chemical reactions. We have considered the edges of 
metabolic networks as chemical reactions by following the notation from the KEGG database. The chemical compounds are also called
substrates. Each substrate can be represented as a node of the graph, linked by chemical reactions. The products 
of these chemical reactions appear as other nodes (substrates). The enzymes are chemical entities which catalyze the chemical reactions. In Fig. 1 
we show the Enzyme Commission \cite{enzyme}
numbers [{\it EC a.b.c.d}] for several enzymes as an example. One enzyme can catalyze more than one chemical reaction. In our work 
we only focus on the chemical reactions, and they are represented by the edges in the metabolic network.}.
In this 
example, nodes and edges 
have physical meaning. There are five
chemical compounds and four chemical reactions in that representation. 
>From this representation we can construct the graph $G$ which only contains compounds as nodes (chemical compound network). Two nodes
are linked by an edge when they occur (either as substrates or products) in the same chemical reaction. After applying the line graph 
transformation on the graph $G$, we obtain the graph $L(G)$ which only contains reactions as nodes (reaction
network). The graph $G$ (compound network) will
have a degree
distribution $P_G(k)$, and the graph $L(G)$ (reaction network), obtained through the line graph transformation,
will allow us to study its degree distribution $P_{L(G)}(k)$ (or other topological observable). In
particular, we focus on the correlation between both degree distributions $P_G(k)$ and $P_{L(G)}(k)$, through the line graph 
transformation. In 
these figures, we 
see clearly
the close relationship between the line graph transformation and real metabolic networks illustrating the motivation of our analysis. In 
this example, the line graph transformation becomes more meaningful. 


However, we must notice that in some cases the line graph transformation could 
give rise to spurious nodes. To be concrete, we will use the terminology 
of chemical compounds/reactions, but this explanation can be extended to any other similar graph structure.
These spurious nodes 
appear when the reactions have more than one product (or substrate) as chemical compounds. When these cases appear, the compound graph $G$
generated (as explained later with Fig. 6) will have extra edges, and consequently the line graph transformation 
will generate reactions as nodes which do not exist in the real metabolic process. We have analysed this issue and a detailed
explanation for the case of metabolic networks can be found in Sec. III.

\begin{figure}[h]
\centerline{\protect
\hbox{
\psfig{file=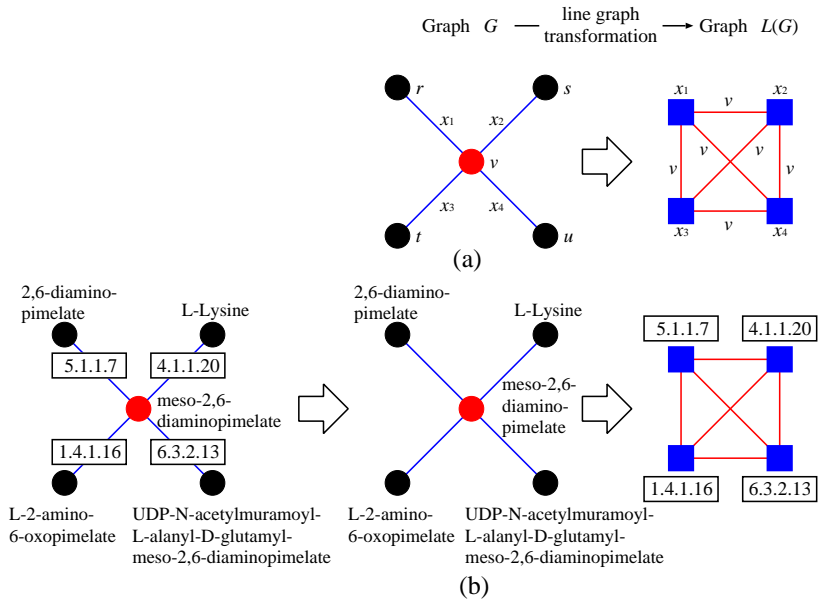,height=8.0cm,angle=0}}}
\caption{\small{(a) Graph $G$ is an initial network. $L(G)$ is the corresponding line graph network. 
(b) A representation of one subgraph from the Lysine Biosynthesis metabolic network. The graph $G$ only contains 
compounds as nodes (chemical compound network). Two compounds are linked by an edge when they occur in the same chemical reaction.
Applying the line graph transformation we generate the line graph $L(G)$, which is the chemical reaction graph 
embedded in the metabolic network, where the nodes are the reactions. This example illustrates the close 
relationship between the line graph transformation and the metabolic network.}}
\end{figure}



The main feature underlying the line graph transformation can be summarized as follows: 
We assume that the initial network has the scale-free topology as $P(k)\approx k^{-\gamma}$. 
As the degree of each transformed node (i.e., an edge in $G$) will be roughly around $k$, the distribution of the line graph $L(G)$
should be $k\cdot k^{-\gamma}$ = $k^{-\gamma + 1}$ with degree around $k$. Therefore, we can conclude that if we have a 
graph $G$ with a probability distribution following a power-law as $k^{-\gamma}$, then
$L(G)$ will follow a power-law as $k^{-\gamma + 1}$.
We have developed two models to reproduce the behavior of the line graph transformation over the scale-free network. In the 
first one we solve 
the discrete equation for the degree distribution of the line graph and in the second one we use the inverted beta distribution.
A detailed mathematical explanation on these models can be found in the next subsections.

\begin{figure}[h]
\centerline{\protect
\hbox{
\psfig{file=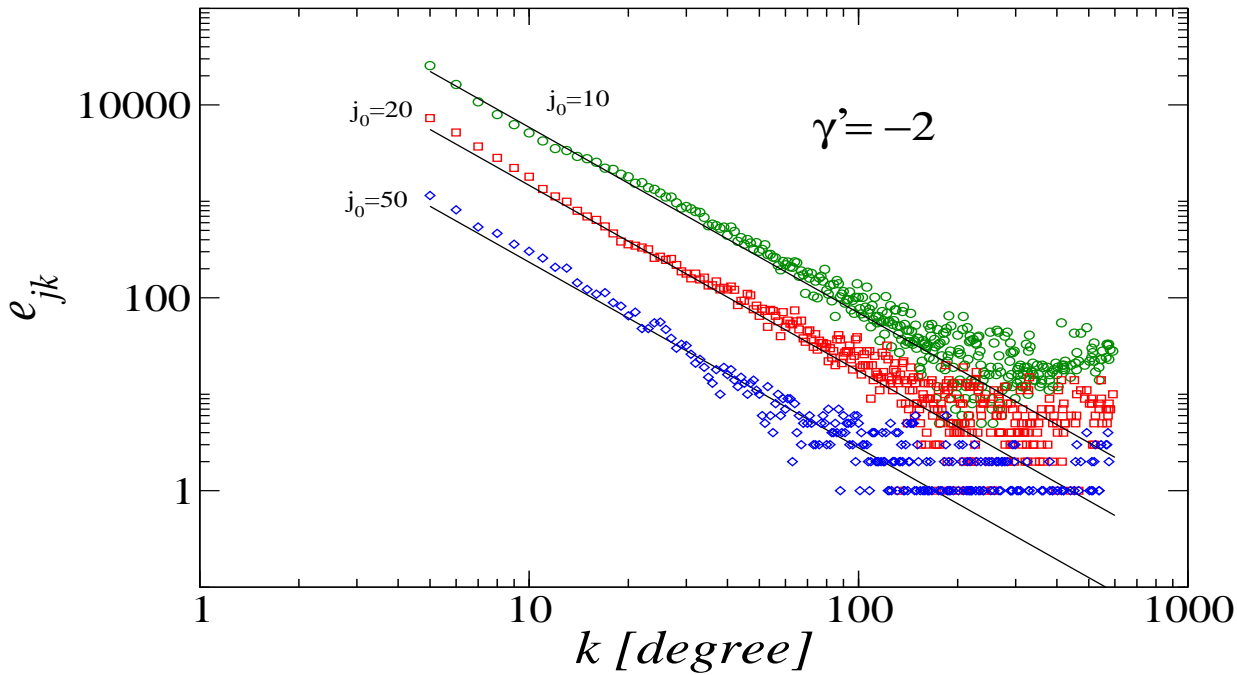,height=11.0cm,width=17.0cm,angle=-90}}}
\caption{\small{Eq. (3) is numerically computed with the BA model by fixing
degree $j=j_0$ ($j_0=10,20,50$). 30 trials of BA networks 
of 10000 nodes are computed for each $j_0$ in order to have 
enough statistics. The results reproduce Eq. (4) with
$\gamma^\prime=-2$ (continuous lines).}}
\end{figure}

\subsection{Discrete equation for the degree distribution of a transformed graph}

We asumme the following: $A)$ degree distribution of an original scale-free network follows $|\{v|deg(v)=d\}| \propto d^{-\gamma}$
if $(d > m_0)$, and $|\{v|deg(v)=d\}| =0$ otherwise. $B)$ edges are randomly generated under $(A)$. Precisely, $deg(u)$ is independent 
of $deg(v)$ for each edge $(u, v)$. Hence, if $deg(u)$ is $1 + d_1$ and $deg(v)$ is $1 + d_2$, the transformed node corresponding to edge $(u, v)$ has degree
$d_1 + d_2$. The following relations hold from $(A)$ and $(B)$: $Prob\,[deg(v)= 1 + d_i] \propto  \,\,\, (1+d_i)^{-\gamma}$ with $i$=1,2,
for a randomly generated edge ($u$, $v$), where $d_1+1$, $d_2+1$ $\geq$ $m_0$. Therefore, degree distribution of a transformed node 
would be given by:

\begin{equation}
Prob\,[deg{(u,v)}=d] \propto \sum_{d=d_1+d_2 \atop {d_1+1\geq m_0 \atop {d_2+1\geq m_0}}} (1+d_1)^{-\gamma + 1} \cdot (1+d_2)^{-\gamma + 1}.
\end{equation}

This equation can be expressed as:

\begin{equation}
Prob\, [deg{(u,v)}=d]\propto \sum_{x=m_0}^{d_0-m_0} [{x^{1 -\gamma}\cdot (d_0-x)^{1 -\gamma}}] 
\end{equation}
with $d_0$ = $d + 2$ and $x= 1 +d_1$.
We are also able to sum in a discrete way  in terms of polygamma functions (see the next subsection).

Though we have assumed that $deg(u)$ is independent from $deg(v)$, this assumption can be weakened to the condition
that the original network shows no assortative mixing (neutral network). By assortative (disassortative) 
mixing in networks we understand the 
preference for nodes with high degree to connect to other 
high (low) degree nodes \cite{new_newman, newman2}.  
We could choose randomly one edge and we consider the node reached by following that edge.  
Following {\it Newman} \cite{new_newman}, the degree distribution for the node at the end of a randomly chosen
edge will be: $(k+1)\cdot P(k+1)$,  if we only consider the number of edges leaving the node (i.e., not taking into account 
 the node we arrived along). Therefore, the probability distribution $e_{jk}$ of all edges ($u$,$v$) that link together nodes with 
 degree $j+k$ (sum of the degrees of the nodes at the ends of $(u,v)$ edge)
 would be approximated as:

\begin{equation}
e_{jk} \propto (j+1)\cdot (j+1)^{-\gamma} (k+1)\cdot (k+1)^{-\gamma}.
\end{equation}

In order to validate Eq. (3) with the BA model, we compute $e_{jk}$  by using several fixed values of $j=j_0$. By fixing $j$, Eq. (3) can be 
written as: 

\begin{equation}
e_{{j_0k}}\propto C(j_0)\cdot(k+1)^{-\gamma+1},
\end{equation} 
where $C(j_0)$ is a constant, and as $\gamma=3$ for the BA model, $\gamma^\prime=-\gamma+1$ should be $-2$. We show the results of 
our numerical computation in Fig. 2. We see three power-law with exponent
$\gamma^\prime=-2$ for three different values of $j_0$. These results indicate that Eq. (3) 
is a valid expression for the BA model. It is worth noticing that $e_{jk}=e_{kj}$.

\begin{figure}[h]
\centerline{\protect
\hbox{
\psfig{file=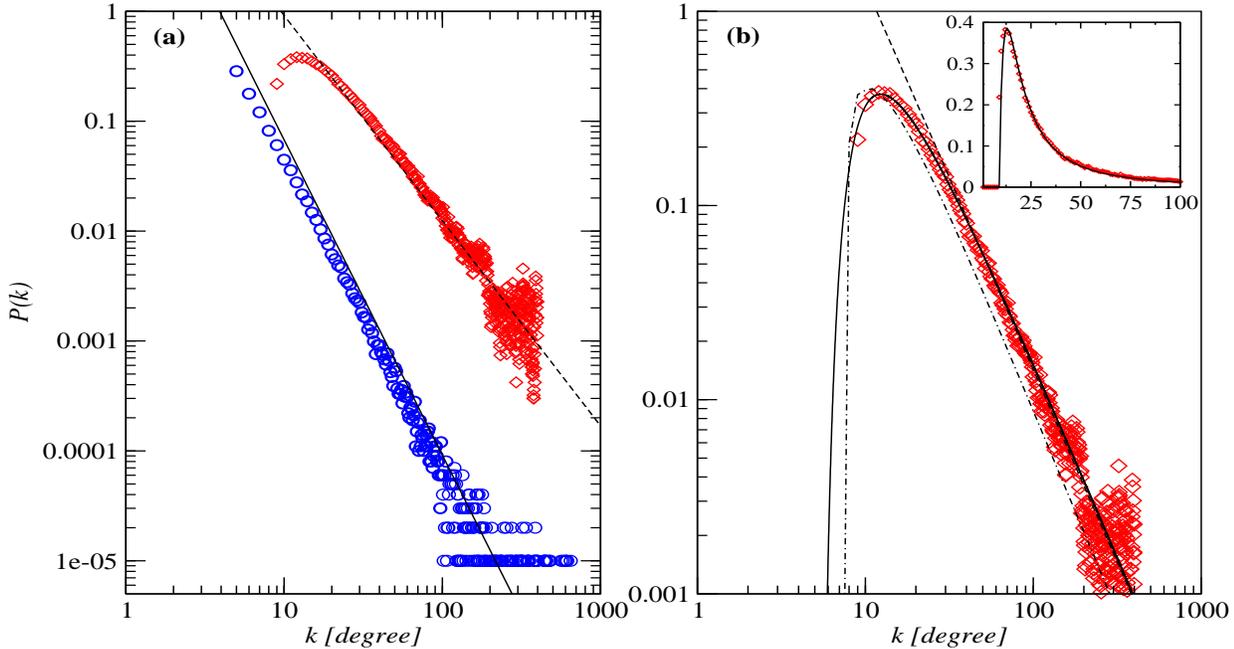,height=11.0cm,width=17.0cm,angle=-90}}}
\caption{\small{Analysis with synthetic network. (a) Circles (blue): degree distribution
generated by using the {\it Barabasi-Albert} model \cite{7, 10} with $m_0$=$m$=5 (in what follows, $m_0$=$m$). The synthetic 
network follows 
the power-law distribution $P(k)\approx k^{-\gamma}$ with exponent $\gamma$ = $2.9\pm 0.1$ (continuous line). Diamonds (red): the 
distribution 
of the transformed scale-free network also follows a power-law with exponent
$\gamma$= $1.9\pm 0.1$ (dashed line). We took average of 10 BA networks of
10000 nodes.(b) Inverted beta distribution (continuous line) with parameters $\beta=17$,
$\alpha=1$ and $a=m_0=5$ and polygamma distribution for $\gamma=3$ (dash-dotted line). In the inset we show the transformed network data
together the polygamma distribution in linear-linear scale. It is interesting to note that the peak 
found in the transformed network does not exist in the original network. This feature can also be predicted by our theoretical model.}}
\end{figure}

In order to compare with Eq. (1), we sum for all the possible degrees of the two vertices at either end of 
edges ($u$,$v$) which can generate
transformed nodes with degree $d$:

\begin{equation}
Prob\,[deg{(u,v)}=d] = \sum_{d=j+k \atop {j+1\geq m_0 \atop {k+1\geq m_0}}} e_{jk} \propto 
\sum_{d=j+k \atop {j+1\geq m_0 \atop {k+1\geq m_0}}} (j+1)^{-\gamma+1} \cdot (k+1)^{-\gamma+1},
\end{equation}
and Eq. (1) holds.

Although in \cite{new_newman} it is suggested that the BA model is not assortative (neutral scale-free network)
(i.e., Pearson correlation function gives a zero result), {\it Redner} and {\it Krapivsky} \cite{redner} pointed 
out that obvious degree correlations exist in BA model (older nodes tend to be connected) and in similar growth models.
However, as we have seen in Fig. 2, Eq. (3) is a good approximation for the BA model.



\subsection{Analysis in terms of Polygamma functions}

The digamma function is defined as:
$\Psi(z)=\frac{d}{dz}\ln\Gamma(z)=\frac{\Gamma^{\prime}(z)}{\Gamma(z)}$
where $\Gamma(z)$ is an extension of the factorial to complex and real
numbers and it is analytic everywhere except at $z=0,-1,-2,-3...$. The $n$th derivate of $\Psi$ is named the Polygamma function, denoted 
$\psi_n(z)$. The notation $\psi_0(z)$ is frequently used for the digamma
function itself as $\psi_n(z)=\frac{d^n}{dz^n}\frac{\Gamma^{\prime}(z)}{\Gamma(z)}=
\frac{d^n}{dz^n}\psi_0(z)$. Other useful identities are
\begin{equation}
\frac{d\psi_0}{dz}=\sum_{k=0}^{\infty} \frac{1}{(z+k)^2} 
\end{equation}
and

\begin{equation}
\psi_0(x)=-\gamma + \sum_{k=1}^{x-1} \frac{1}{k}
\end{equation}
where $\gamma$ is the Euler-Mascheroni constant and $x$ is a positive integer value.
 

By decomposing Eq. (2) in partial fractions, and by using the polygamma relationships shown above, we can 
find the expressions for Eq. (2) evaluated with $\gamma=2$ and $\gamma=3$. We detail the calculation for $\gamma=2$.
The summation of Eq. (2) can be decomposed in two terms:

\begin{equation}
\sum_{x=m_0}^{d_0-m_0} \frac{A}{x} + \frac{B}{d_0-x}
\end{equation}
where $A=B=1/d_0$.

The first term of the sum gives: 
\begin{equation}
\sum_{x=m_0}^{d_0-m_0} \frac{1}{x}= \frac{1}{m_0} + \frac{1}{m_0+1} + ...+\frac{1}{d_0-m_0+1}+\frac{1}{d_0-m_0}
\end{equation}
and by using Eq. (7), we can write it as: $\psi_0(x+1)- \psi_0(m_0)$, with $x=d_0-m_0$. Analogously, the second term 
gives the same contribution.
Therefore for $\gamma=2$, Eq. (2) reads as:

\begin{equation}
Prob\,[deg{(u,v)}=d]_{\gamma=2} \propto \frac{2}{d_0}[\psi_0(x+1)- \psi_0(m_0)].
\end{equation}
For $\gamma=3$ we have 

\begin{equation}
Prob\,[deg{(u,v)}=d]_{\gamma=3} \propto \frac{2}{d_0\, ^3}[\psi_0] + \frac{1}{d_0\, ^2}[\psi_1]
\end{equation}
with 

\begin{equation}
\psi_0= 2[\psi_0(x+1)- \psi_0(m_0)] 
\end{equation}
and 

\begin{equation}
\psi_1= 2[\psi_1(m_0)-\psi_1(x+1)], 
\end{equation}
where $d_0=d+2$ and $x=d_0-m_0$ are integer values.

It is interesting to note that $Prob\,[deg{(u,v)}=d]_{\gamma=2} \approx d^{-1}$ and 
$Prob\,[deg{(u,v)}=d]_{\gamma=3} \approx d^{-2}$ for large $d$, which
matches the distribution of $k^{-\gamma + 1}$ as we can see in Fig. 3(b).


\subsection{Inverted beta distribution}

One drawback of the previous approach is that the exponent $\gamma$ is considered as an integer number which constrains its range of 
quantitative applicability. Hence, we have looked for a continuous function in terms of the exponent $\gamma$. In that sense, we have found 
that the inverted beta distribution $B(y) \propto {(y-a)^{\beta-1}}/{(1+y-a)^{\alpha + \beta}}$, which is obtained doing the 
transformation  $Y=\frac{1-X}{X}$ over the beta distribution followed by a translation $y\rightarrow\ y-a$, fits well our requirements
and it was used successfully
to reproduce the data. In addition, it is interesting to note that the distribution shows a power-law tail
for large $y$ as: $B(y)\rightarrow {y^{-\alpha-1}}$.


\subsection{Theoretical results}

Once the theoretical approaches have been introduced, we generate a synthetic scale-free network with exponent $\gamma = 2.9$  
using the BA
model \cite{7,10} and we study the behavior of the line graph transformation over that scale-free network. We  
compare the transformed network with the results from the theoretical functions introduced in the previous sections. In 
Fig. 3(a) we see that the transformed network (diamonds) follows a power-law with exponent $\gamma=1.9$. It is interesting 
to note that the exponent of this scale-free is exactly reduced by one unit (from the original scale-free network to the transformed network)
 as it was predicted in the previous sections. As a second result, a peak was found 
for low degree nodes in the line graph transformed network, indicating that the power-law
is like a tail (or asymptotic behaviour) of a more general distribution. In Fig. 3(b) we plot the inverted beta distribution 
and the polygamma function to compare with the transformed network. We see that the curves reproduce well the peak of the data for low 
degree $k$ and also shows a power-law tail for higher degree $k$. 
Both agreements give us confidence about the fairness of both approaches used to study the data.


\section{Experimental data}

There are several examples, in biological and non-biological networks, which appear to support our theoretical analysis. We have tested 
our results in the World Wide Web network with a size of 325729 nodes representing web pages being connected by
links each other. The dataset was obtained from the website of {\it Notre-Dame Research Group} \cite{20}. We must notice 
that we have considered the links as undirected edges in order to compare with our theoretical approach. 
In that sense the value of $\gamma$ obtained here could be considered as an average of the $\gamma_{in}$, $\gamma_{out}$ \cite{21}. We have
also analysed the protein-protein interaction network for the yeast {\it S. Cerevisae} which contains around 1870 proteins 
as nodes linked by 2240
bidirectional interactions \cite{9, 20}. In Fig. 4 and Fig. 5 we show the data for the {\it WWW} network and the protein-protein interaction network respectively. 
We see that both networks (circles) are following the power-law $P(k)\sim k^{-\gamma}$. In the same figures, we present our results for the corresponding transformed
network (diamonds). In both cases, we see that the scale-free topology is preserved and the exponent $\gamma$ is decreased by almost one unit as we expected.

\begin{figure}[h]
\centerline{\protect
\hbox{
\psfig{file=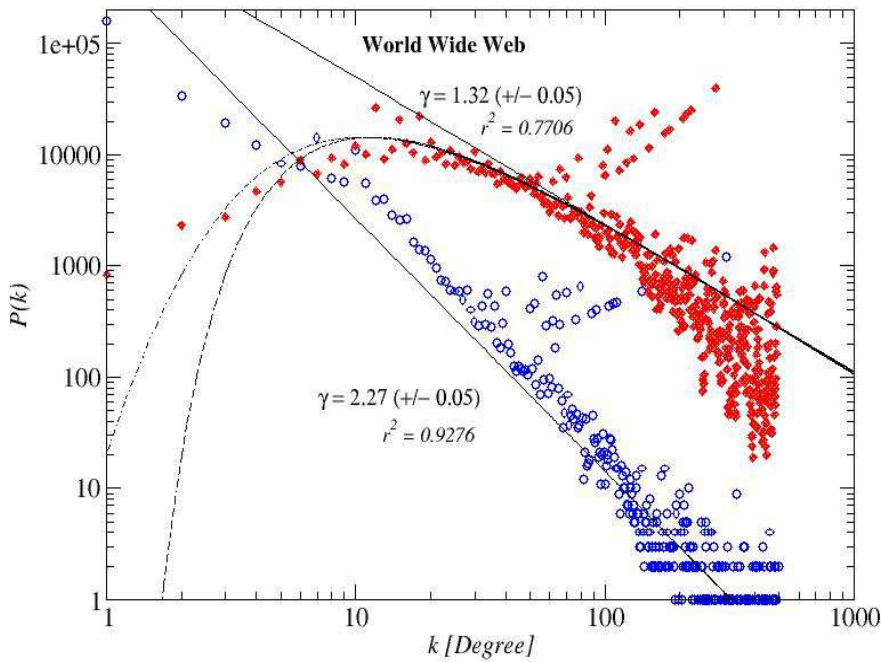,height=10.0cm,width=12.0cm,angle=0}}}
\caption{\small{Circles (blue): degree distribution for the {\it WWW} network generated with data 
from \cite{20}. The {\it WWW} network follows a power-law with exponent $\gamma=2.27$. Diamonds (red): the  
distribution of the transformed network also shows a power-law tail with exponent $\gamma=1.32$. We show 
the inverted beta distribution evaluated with $a=m_0=1$ (dashed line), $a=m_0=0$ (dash-dotted line), and the beta parameters 
are $\beta=16$, $\alpha=0.4$. The correlation coefficient is $r^2$.}}
\end{figure}

\begin{figure}[h]
\centerline{\protect
\hbox{
\psfig{file=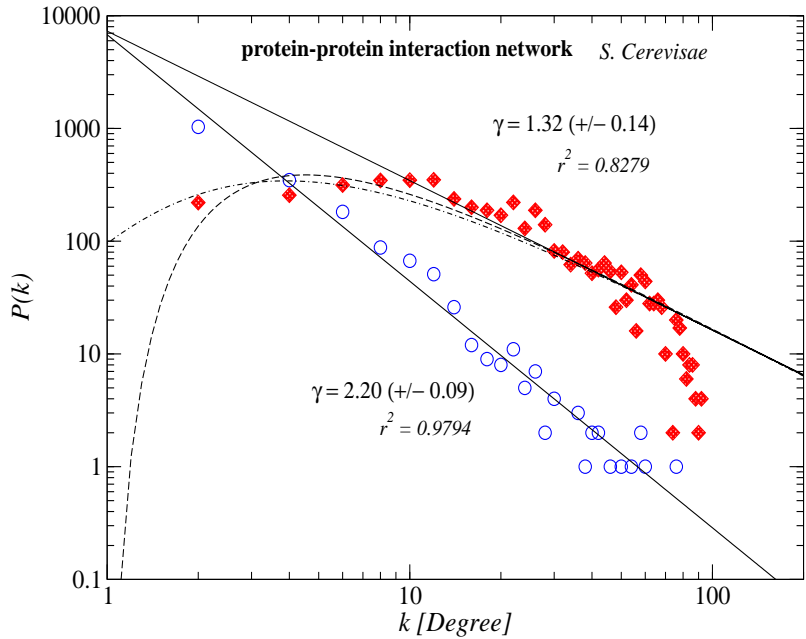,height=10.0cm,width=12.0cm,angle=-90}}}
\caption{\small{Circles (blue): degree distribution for the 
protein-protein interaction network {\it S. Cerevisae} generated with data  
from \cite{20}. The protein-protein interaction network follows a power-law with exponent $\gamma=2.20$. 
Diamonds (red): the distribution of the transformed network 
also shows a power-law tail with exponent $\gamma=1.32$. We show 
the inverted beta distribution evaluated with $a=m_0=1$ (dashed line), $a=m_0=0$ (dash-dotted line), and the 
beta parameters are $\beta=6$, $\alpha=0.4$. The correlation coefficient is $r^2$.}}
\end{figure}

Following with our analysis of real networks, we analyse now the metabolic network from the KEGG database \cite{19} which contained
around 10400 chemical compounds and 4100 chemical reactions. The KEGG database is one of the best sites 
for biochemistry, metabolism, and molecular biology information. As we showed in Fig. 1, the line graph transformation is closely
related to the representation of metabolic networks, however we must point out some differences which we have found in our study.

\begin{figure}[h]
\centerline{\protect
\hbox{
\psfig{file=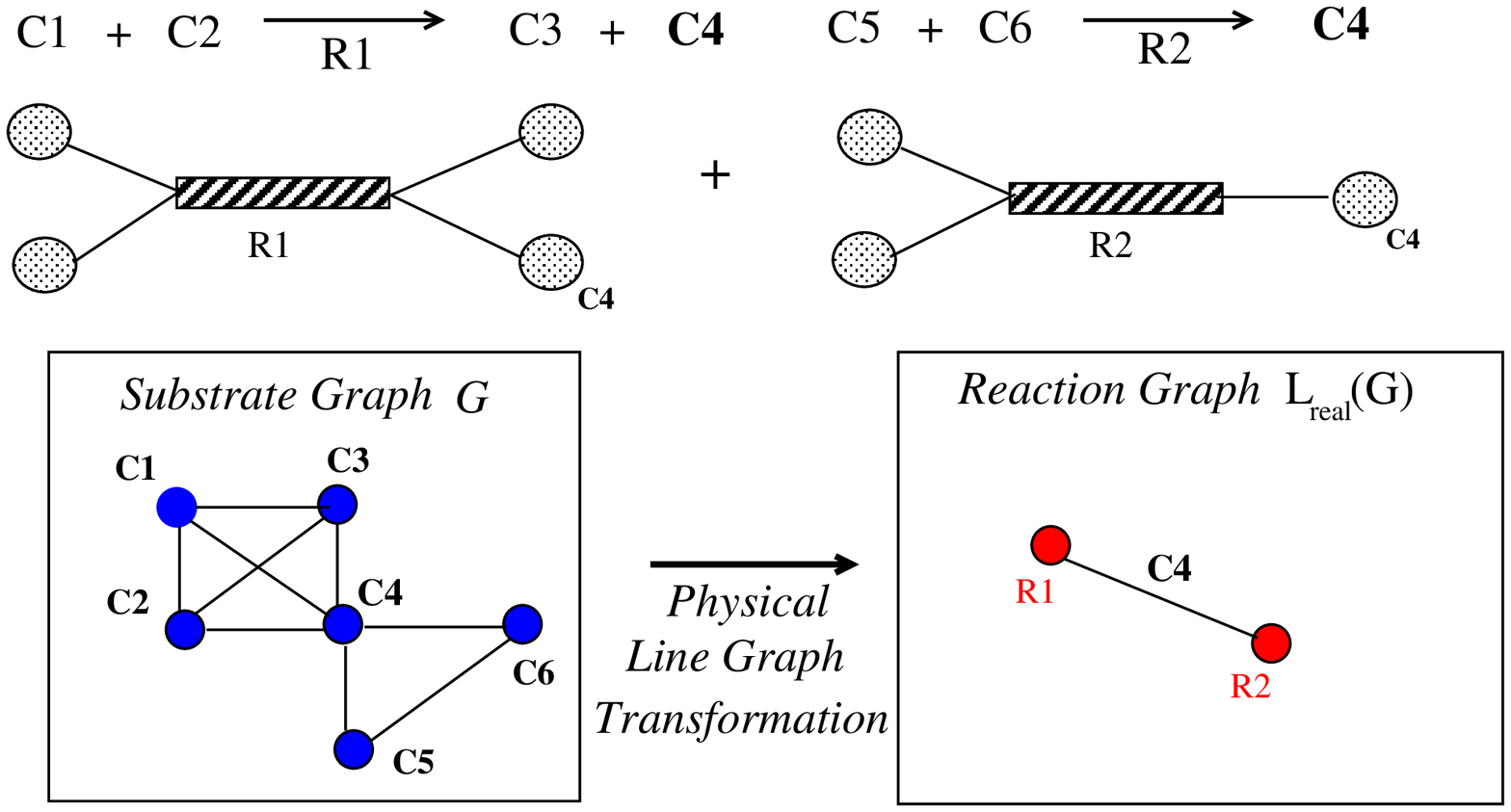,height=10.0cm,width=12.0cm,angle=-90}}}
\caption{\small{Sketch of metabolic networks. We show two reactions ($R1$, $R2$), with only one common substrate (product) $C4$.
The substrate graph $G$ (compounds) is shown with dark blue circles. 
The reaction graph $L_{real}(G)$ (reactions) is shown with light red circles. To obtain this graph $L_{real}(G)$ we apply a ${\it physical}$ 
line graph transformation, i.e., we do not transform all the edges from the initial graph and we only generate 
the same number of nodes in the
transformed network as the number of reactions in the real metabolic process.
An ordinary line graph transformation would generate
nine reactions as nodes in the line graph $L(G)$, when only two reactions exist in the real network.}}
\end{figure}

As it is depicted in Fig. 6, in the metabolic networks exist some cases where not all the edges associated to the substrate network
should be transformed by the line graph transformation. The main issue 
is that for each chemical reaction with more than one product (or substrate), we must only transform the 
same number of edges from the substrate 
graph as the number of reactions in the real metabolic process. In the case shown in Fig. 6, the resulting 
network (reaction network), after an ordinary line graph transformation, would generate up to nine nodes (reactions), but 
only two exist in the
real metabolic process. These spurious nodes only appear when one (or some) reaction(s) in the network have 
more than one product (or substrate).
These cases should be transformed as it is shown in Fig. 6 (i.e., by creating only as many nodes 
in the transformed network as reactions
 in the real metabolic process). This procedure is called ${physical}$ line graph transformation.
The graph generated by the 
${physical}$ line graph transformation is denoted by $L_{real}(G)$.
In a network where the are some spurious cases as exposed above, we proceed by applying the $physical$ line transformation for these ones,
and the ordinary line graph transformation for the rest of the cases.
The resulting distribution could be different if we 
compare it with the distribution generated only by
the line graph
transformation
 $P(k)$ $\simeq$ $k^{-\gamma + 1}$. The distortion could be larger if there are
many of these spurious cases in a network.

In Fig. 7 we show the degree distribution of the
chemical compounds (circles) in the metabolic network from the KEGG database. In the same figure, we 
have plotted the distribution (diamonds) for the transformed network which corresponds to the 
reaction network.
We see that both distributions follow a power-law and the
difference between their exponents is one. However, that transformed network $L(G)$ would not have
a full physical meaning because it could have extra reactions as nodes which do not exist in the real metabolic process,
as we explained in Fig. 6. In Fig. 8 we plot the transformed network
after applying the line graph transformation and the $physical$ line 
graph transformation for the spurious cases. We
see that both graphs are scale-free networks, 
but in this case, the difference between the exponents is smaller than one, due to the distortion mentioned above. 

Experimental results shown in our paper suggest that exponents of the transformed scale-free network 
may change due to the effect of the spurious cases. However, the experiments also indicate that
the scale-free topology is preserved. In particular, if the average number of substrates and products 
per reaction is small, the spurious cases may not strongly affect, and 
the scale-free power-law distribution would be preserved.

\begin{figure}[h]
\centerline{\protect
\hbox{
\psfig{file=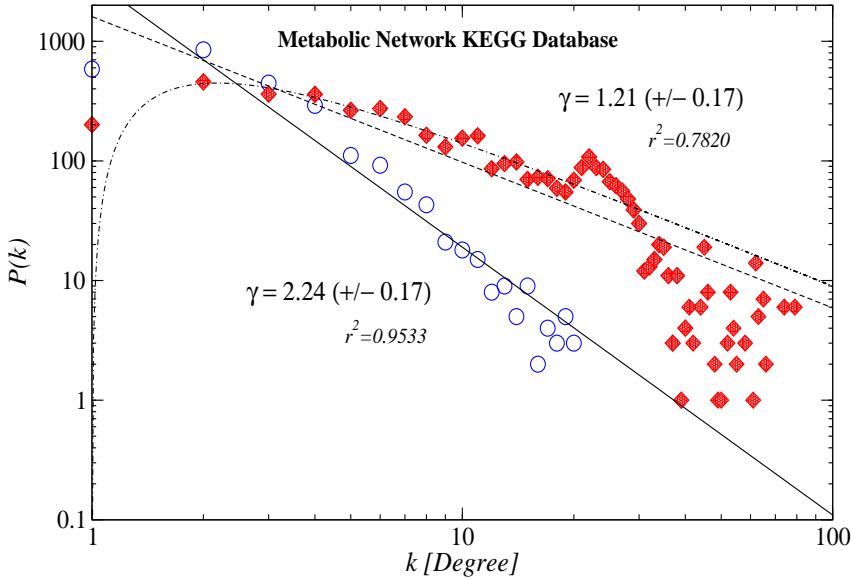,height=10.0cm,width=12.0cm,angle=-90}}}
\caption{\small{Circles (blue): Degree distribution for the metabolic networks of KEGG database \cite{19}. The 
data included the following 9 metabolic pathways: {\it Carbohydrate Metabolism}, {\it Energy Metabolism}, {\it Lipid Metabolism}, {\it Nucleotide Metabolism}, 
{\it Amino Acid Metabolism}, {\it Metabolism of Other Amino Acids}, {\it Metabolism of Complex Lipids}, 
{\it  Metabolism of Complex Carbohydrates} and {\it Metabolism of Cofactors and
Vitamins}. Diamonds (red): The metabolic network follows a power-law with exponent $\gamma=2.24$ (continuous line). 
Diamonds (red): the transformed distribution of the scale-free network shows a power-law tail with exponent $\gamma=1.21$ (dashed line). 
With dash-dotted line, we show the inverted beta distribution with parameters $\beta=2.5$,
$\alpha=0.25$ evaluated with $a=m_0=1$. The correlation coefficient is $r^2$.}}
\end{figure}

\begin{figure}[h]
\centerline{\protect
\hbox{
\psfig{file=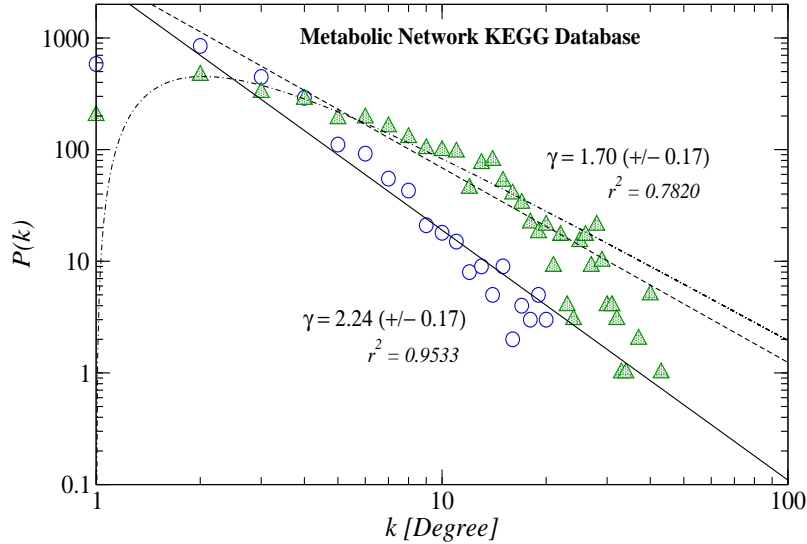,height=10.0cm,width=12.0cm,angle=-90}}}
\caption{\small{Circles (blue): Same as Fig. 5. Triangles (green): the distribution
of the transformed network after applying the line graph transformation and the $physical$ line graph transformation 
for the spurious cases. It shows a power-tail with exponent $\gamma=1.70$ (dashed line). With dash-dotted line, 
we show the inverted beta distribution with 
parameters $\beta=2.7$,
$\alpha=0.7$ evaluated with $a=m_0=1$. The correlation coefficient is $r^2$.}}
\end{figure}


It is interesting to quote a previous work \cite{18}, which analysed the distribution of metabolite 
connectivities in both substrate and reaction graphs. Although in that paper they do not mention about the line graph transformation 
technique
and, consequently, they do not discuss about the reason of the modification of exponent $\gamma$, they notice that the degree distribution
in the reaction graph does not follow a simple power-law and it appears to be governed by two quantitatively different regimes.


\section{Conclusions}

We have reported on the two complementary representations of a scale-free network using the line graph transformation. This 
transformation is useful when it is applied on networks where the edges have physical meaning. In particular, we have 
illustrated that the line graph  transformation is closely related to the representation of metabolic network.
In this network by using the line graph transformation, the reaction network 
can be generated and it allows us to study its degree distribution $P(k)$ and other topological observables as clustering degree, for example.

The two goals of the present 
work can be summarized as follows: We have described the real networks as two complementary representations of a scale-free 
network, where the second one emerges when the line graph transformation
is done over the first one. Our second goal is that we have found that the exponent $\gamma$ is always one unit less than 
the initial exponent $\gamma$ coming from the original scale-free network. We have seen that the 
difference of one unit in the exponents of the degree distributions 
can be found in biological and non-biological networks. We 
have also carried out a 
theoretical study of the general distribution underlying
the line graph transformation, being successfully to reproduce the tail of the power-law and the peak found for low-degree nodes.
However, we should bear 
in mind that in some particular cases (e.g., metabolic networks) an ordinary line graph
transformation applied over a real network could generate extra nodes in the transformed network without a real correspondence. These 
spurious cases make that the difference between the exponents $\gamma$ could differ from one unit. 

We also took advantage of the fact that the BA model is not assortative \cite{new_newman} to compare our results 
for the discrete equation of a transformed graph. As a future work, an extension of this analysis to assortative networks should be done.

This study is an interesting step forward to understanding large complex networks 
from this complementary scale-free perspective. 


\vspace{0.5cm}
\noindent


\noindent
{\bf{Acknowledgements}}

This work 
was partially supported by Grant-in-Aid for Scientific Research on Priority Areas (C) ``Genome Information Science''
from MEXT. We also thank the Research Group of Notre Dame University for making its database
publicly available for research purposes.

\end{document}